\documentstyle[aps,prb,epsfig]{revtex}

\title{Fabrication of Nano-Scale Gaps in Integrated Circuits}

\author{Roman~Krahne, Tali~Dadosh*, Amir~Yacoby, Hadas~Shtrikman, Joseph~Sperling*, and Israel~Bar-Joseph}
\address{Department of Condensed Matter Physics,\\ {*}Department of Organic Chemistry,\\
Weizmann Institute of Science, 76100 Rehovot, Israel}
\begin{document}
\maketitle
%

\begin{abstract}
Nano-size objects like metal clusters present an ideal system for
the study of quantum phenomena and for constructing practical
quantum devices. Integrating these small objects in a macroscopic
circuit is, however, a difficult task.  So far the nanoparticles
have been contacted and addressed by highly sophisticated
techniques which are not suitable for large scale integration in
macroscopic circuits. We present an optical lithography method
that allows for the fabrication of a network of electrodes which
are separated by gaps of controlled nanometer size. The main idea
is to control the gap size with subnanometer precision using a
structure grown by molecular beam epitaxy.
\end{abstract}



Recently, there is growing interest in the study of the electrical
properties of nano-size objects. In these small scale structures
the energy associated  with size quantization and the Coulomb
energy are considerably enhanced and can exceed the thermal energy
$kT$ at room temperature. Therefore, such systems are ideal for
studying quantum phenomena and for constructing practical quantum
devices. The single-electron transistor (SET) has emerged as an
important experimental platform in both aspects. It consists of a
nano-size metal island which is connected through two tunnel
barriers to the leads\cite{fulton87.1}. The SET also presents most
of the problems that characterize device physics at this length
scale: fabrication of macroscopic leads with a nanometer-size gap,
definition of a nano-size island, positioning it between the leads
and controlling the tunnel barriers with high precision. The
fabrication of the leads has proven to be a particularly
challenging task. It requires nano-scale resolution for defining
the gap, and the ability to integrate the device in a highly
connected and complex macroscopic circuit. A variety of solutions
have been suggested to this issue, of which the majority are
appropriate for contacting only a single device. These include
using the tip of a scanning tunnelling microscope for
contacting\cite{porath97.1}, electromigration\cite{park99.1},
breaking thin wires \cite{reed97.1}, or metal deposition on
electron-beam defined
structures\cite{klein96.1,porath00.1,bezryadin97.1}. Other
techniques attempted  to use a conducting DNA-based
network\cite{braun98.1} to solve the connectivity issue. Recently
logic circuits with carbon nanotube have been
reported\cite{bachtold01.1}.

In this paper we present a method that allows us to fabricate a
network of electrodes that are separated by gaps of controlled
nanometer size using conventional optical lithography. The gap
size is determined by the thickness of a molecular beam epitaxy
(MBE) grown layer embedded in a sandwich structure. This layer is
selectively etched and after metal evaporation a gap is formed
exactly where the layer was removed. Controlled positioning of the
nanoparticles between the electrodes is done by electrostatic
trapping\cite{bezryadin97.1}. We show that the device exhibits
clear Coulomb blockade (CB) characteristics in the transport
measurements and can function as an SET.

\begin{figure}
\epsfxsize 8.5cm
\begin{center}
\epsffile{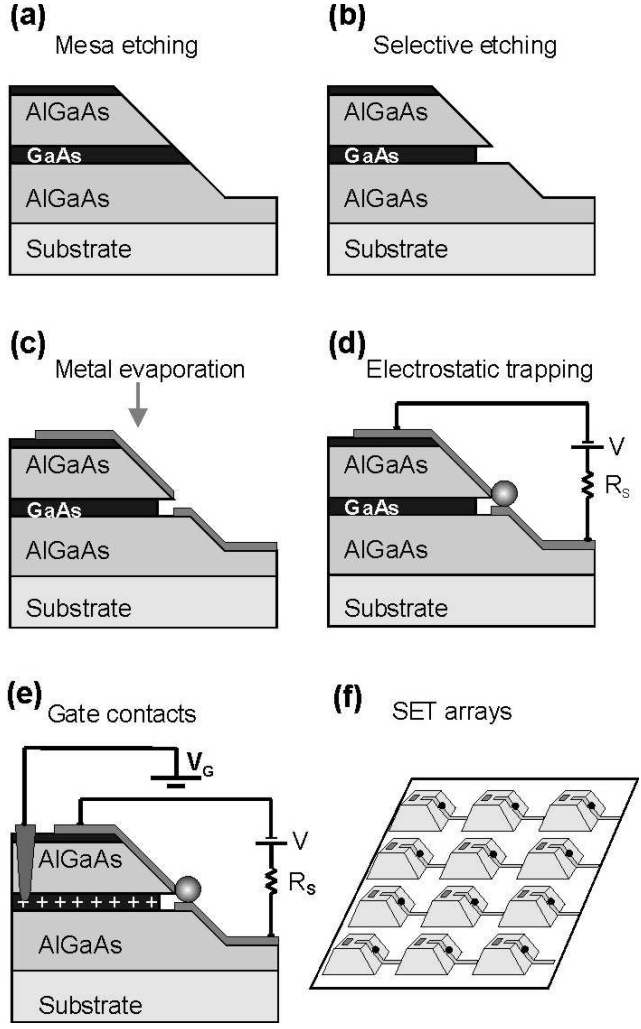}
\end{center}
\caption{(a) Sketched description of the fabrication process: a
mesa structure is defined by optical lithography and wet etching,
(b) selective wet etching removes some tens of nanometers of the
GaAs layer, (c) after metal evaporation two electrodes are
separated by a gap of a few nanometers, (d) the nanocluster is
positioned by electrostatic trapping \cite{bezryadin97.1}, (e) a
heavily doped GaAs layer could be used as a gate, (f) the process
allows for the fabrication of a network of devices with gaps of
controlled nanometer size.} \label{setup}
\end{figure}

The sample fabrication is illustrated in Fig.~\ref{setup}~(a)-(d).
A quantum-well (QW) structure is grown on a GaAs substrate by MBE.
 The thickness of the two undoped AlGaAs layers
above and below the QW are 200 nm and 100 nm, respectively. The
thickness of the embedded GaAs layer is varied between 5 to 10 nm
in various samples. In the first step, 200 nm high mesa structures
are defined by optical lithography and a standard wet-etching
process such that the GaAs layer is exposed on the side of the
mesa (Fig.~\ref{setup}~(a)). Next, the pattern of the electrodes
is defined by optical lithography. On the exposed areas a few tens
of nanometers of the GaAs layer (from the side) are removed by
selective wet-etching \cite{salvo92.1} using citric acid and
H$_2$O$_2$ (5:1), as shown in Fig.~\ref{setup}~(b). The
selectivity of the etching between GaAs and AlGaAs is nominally
100:1, hence the width of the etched layer is the QW width. The
electrodes are fabricated by thermally evaporating a thin film (5
to 15 nm) of PdAu from a direction perpendicular to the plane of
the wafer surface such that a gap is formed exactly where the GaAs
was removed (Fig.~\ref{setup}(c)). The size of the gap is
determined by the crystal structure (which can be controlled with
subnanometer precision), surface roughness of the etched
AlGaAs/GaAs interface due to the selective etching (less than one
nanometer for short etching times), and by the metal evaporation
which can be controlled on the nm scale. We manage to fabricate
gaps which are less than 5 nm wide and estimate the minimum gap
size that can be achieved by our method to be about one to two
nanometers. An important advantage of this method is that since
the mesa structure and the electrode pattern are defined by
optical lithography, it allows the simultaneous fabrication of
many electrodes separated by nano-size gaps on the wafer surface
(see illustration in Fig.~\ref{setup}~(f)).

To position the nanoparticles between the electrodes we use the
method of electrostatic trapping \cite{bezryadin97.1}: the
electrodes are immersed in an aqueous or organic solution of Au
clusters. The Au clusters can be chemically fabricated with
controlled size in the range of 1 to 100 nm
diameter\cite{schmid94.1}. By applying a voltage between the
electrodes a dipole moment is induced in the Au cluster and it is
attracted to the point of strongest electric field, i.e. the gap,
by dipole-dipole interaction. A series resistance, $R_S$ = 1
G$\Omega$, is integrated in the circuit such that when trapping
occurs the voltage drops on that resistor and the device cannot
trap more clusters. In this way only a single cluster is trapped
in the gap. The setup is sketched in Fig.~\ref{setup}~(d). For
trapping, we apply an AC voltage of 8 V at 50 kHz for a few
seconds. AC voltage is used to increase the impedance of the
liquid solvent and to be able to trap also charged particles.
After trapping, the liquid is blown off with $N_2$.

 In order to form tunnel barriers between the electrodes and
the clusters, the electrodes can be covered with thiol molecules
before the trapping\cite{park99.1,klein96.1}. However, we find
that if the cluster diameter is only slightly larger than the gap,
a tunnel barrier can exist even for uncoated electrodes. This
tunnel barrier could arise from organic molecules on the cluster
surface that stabilize the Au clusters, or from contamination on
the electrodes formed during the trapping process. We note,
however, that the strength of this barrier is reasonably
reproducible as evidenced by transport measurements. The geometry
sketched in Fig.~\ref{setup}~(d) suggests that the contact of the
trapped cluster to the lower electrode is better than to the upper
one. This asymmetry in the contacts should lead to different
tunnel resistances, which is indeed the case as we will see below.

\begin{figure}
\epsfxsize 7cm
\begin{center}
\epsffile{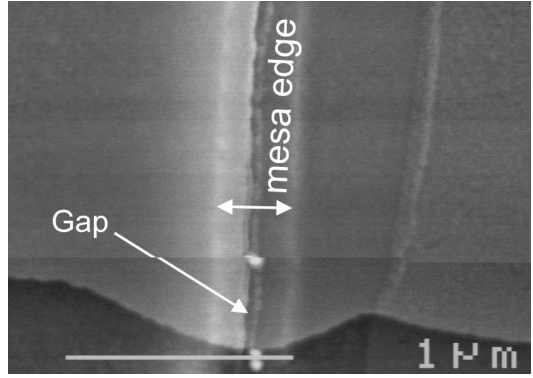}
\end{center}
\caption{SEM image of the metal electrodes separated by a gap
slightly larger than 10 nm and a single trapped Au cluster with 60
nm diameter. Bright regions are covered with PdAu.} \label{sem}
\end{figure}

A scanning electron microscope (SEM) image of a single trapped
cluster between two metal electrodes is displayed in
Fig.~\ref{sem}. In the center of the image we see the mesa edge,
with
 the mesa plateau to the left, and the etched grove separating the
electrodes in the vertical direction. The trapped cluster is seen
below the center of the image. Two more clusters, which do not
bridge the gap, can be observed at the right side of the scale
bar. The gap size is slightly larger than 10 nm and the width of
the electrodes is a few microns each. To enhance the visibility of
the trapped cluster, clusters having a relatively large diameter
of 60 nm are displayed.

\begin{figure}
\epsfxsize 13cm
\begin{center}
\epsffile{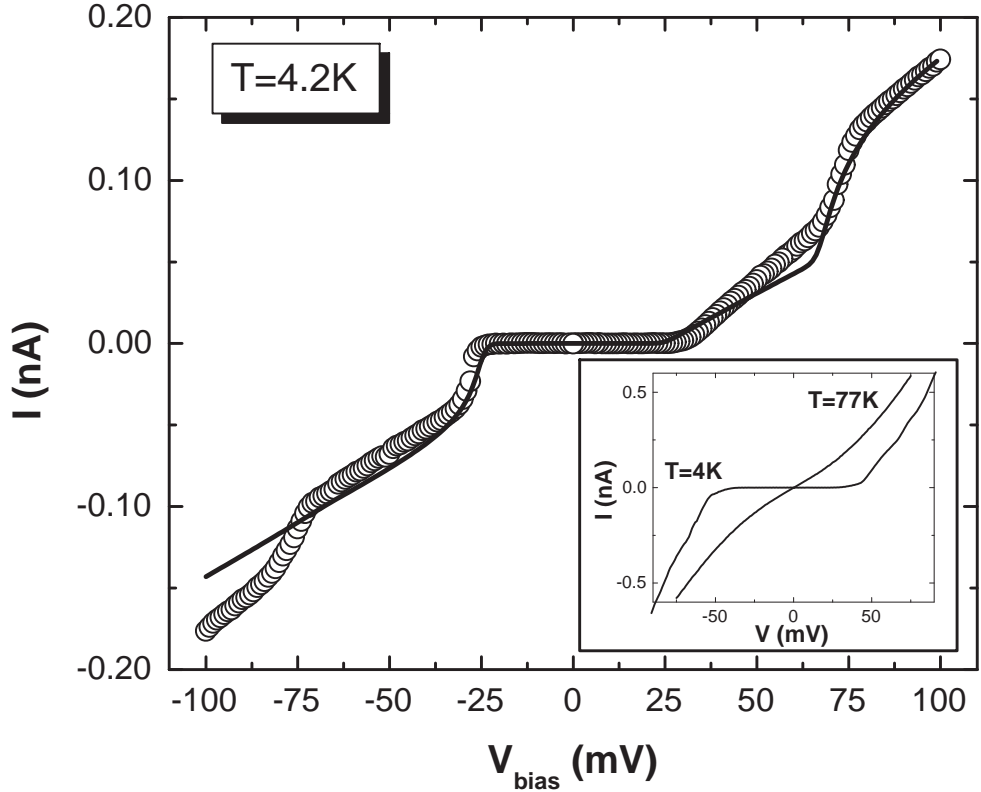}
\end{center}
\caption{Experimental and theoretical $I-V$ curves of trapped
metal clusters. Large circles show experimental data measured on a
Au cluster with 10 nm diameter at 4.2K. The asymmetry with respect
to positive or negative $V_{bias}$, i.e. a linear onset at +25 mV
and a step like onset at -25 mV, results from an asymmetry in
tunnel resistances $R_1$ and $R_2$ to the electrodes and a
residual fractional charge $Q_0$ on the trapped cluster. A
theoretical $I-V$ curve is depicted by the solid line with
fractional charge $Q_0=0.24e$, $R_1=30 $M$\Omega$, $R_2=400
$M$\Omega$, $C_1 =C_2 = 5$~x~$10^{-19}$~F and a stray capacitance
to the electrodes of $C_g=2.5$~x~$10^{-18}$~F. The inset shows
experimental $I-V$ curves for different temperatures of another
sample where a Au cluster with 10 nm diameter was trapped.}
\label{data}
\end{figure}

 Figure \ref{data} shows an $I-V$ curve measured on a
trapped Au clusters with 10 nm diameter at 4.2 K. A clear CB
behavior can be observed up to $V_{bias}\approx \pm 25$~mV. For
larger voltages, $|V_{bias}|>25$~mV, we observe a strong asymmetry
in the $I-V$ curve with respect to positive and negative voltage.
For positive bias voltage we find a linear rise in current in the
range of 25~mV$<V_{bias}<$65~mV, whereas for negative bias we
observe a step-like feature at $V_{bias}=-25$~mV, which then
saturates in a linear slope. This asymmetric behavior has been
discussed by Hanna and Tinkham \cite{hanna91.1} assuming different
tunnel resistances to the electrodes, i.e. $R_2>>R_1$. They
distinguished four different cases in a $C_1/C_2-Q_0$ phase
diagram, each with a qualitatively different onset in conduction.
Here $C_1$ and $C_2$ are the capacitances of the trapped cluster
to the electrodes and $Q_0$ is the residual fractional charge on
the cluster. The solid line displayed in Fig.~\ref{data}
corresponds to the case of $C_1/C_2\approx 1$ and $Q_0=0.24e$. In
order to fit our data \cite{korotkov91.1,korotkov02.1}, we used
$R_1=30$~M$\Omega$, $R_2=400$~M$\Omega$ (which satisfies very well
the assumption of $R_2>>R_1$,) and for the capacitances
$C_1=C_2\approx 5$~x~$10^{-19}$~F, and a stray capacitance to the
electrodes of $C_g=2.5$~x~$10^{-18}$~F.
 The inset in Fig.~\ref{data} shows transport measurements of
another sample at different temperatures, where a Au cluster with
a diameter of 10 nm was trapped. At $T=4.2$~K we find the CB
region to be slightly larger than in the previous sample. Clearly,
also at higher temperatures $T=77$~K a non-linear behavior in the
$I-V$ curve can be observed. Measuring the $I-V$ curve of a
certain device at different times after the cooling reveals the
abrupt occurrence of changes in the characteristics of the
conduction onset and slight changes in the size of the Coulomb
blockade region. This behavior was also observed by Klein and
coworkers \cite{klein96.1} and was attributed to changes in the
local charge distribution in the vicinity of the cluster. Repeated
warming and cooling of the device leads to similar effects.

An important question is how to apply a gate voltage to the
trapped cluster in order to get true transistor functionality.
This is not trivial and to the best of our knowledge, so far no
gated device using metal clusters of nanometer size has been
realized. The major problem is the screening  of the gate voltage
by the comparatively large electrodes in close vicinity to the
small cluster. Our design provides a very elegant method to
position a gate electrode only a few tens of nanometers away from
the cluster. As illustrated in Fig.~\ref{setup}~(e), the GaAs
layer, which has been used for the selective etching, can also be
heavily doped during the MBE growth and then used as a gate. The
great advantage of this method is that due to the mesa design,
each cluster can be gated individually. This would allow large
scale integration of SETs with true transistor functionality on
semiconductor surfaces by optical lithography
(Fig.~\ref{setup}~(f)), especially if the series resistor $R_S$ is
also integrated in the electric circuit on the wafer. This would
enable simultaneous trapping on a large number of devices. Another
approach to position a gate electrode in very close proximity to
the cluster is to grow the lower AlGaAs layer only a few tens of
nanometers thick and to use a n-doped GaAs substrate. In this
design the substrate could serve as a gate, however, this setup
would not allow the individual gating of each cluster. Experiments
to realize a gated structure are in progress and will be reported
elsewhere.

In conclusion we presented a fabrication method to define
compatible contacts to nanosize objects by optical lithography and
selective etching. In transport measurements at low temperatures
we demonstrated Coulomb blockade behavior. In order to operate the
nanoclusters as SETs we proposed a gate design that allows large
scale integration of such devices on semiconductor surfaces.

We want to thank Diana Mahalu for helpful discussions and her
assistance in the early stages of this work. The research of R.~K.
was supported by the "Fritz Thyssen Stiftung" and the European
Commission under contract HPRI-CT-1999-00069. This work was
partially supported by a grant from the Israel Science Foundation
to J.~S.

\bibliographystyle{prsty}

\end{document}